\documentclass[aps,prl,twocolumn,reprint,amsmath,amssymb,superscriptaddress,floatfix,footinbib,longbibliography]{revtex4-1}

\usepackage{graphicx}
\usepackage{epstopdf}

\usepackage[T1]{fontenc}
\usepackage[applemac]{inputenc}
\usepackage{lmodern}
\usepackage[english]{babel}

\usepackage{ae}
\usepackage{units}

\usepackage{amsmath,amssymb,natbib,bm}
\usepackage{psfrag}
\usepackage{subfigure}
\usepackage{amsthm}

\usepackage{fixmath}
\usepackage{booktabs}

\usepackage{slashed}

\usepackage[americaninductors]{circuitikz}
\usepackage{tikz}
\usetikzlibrary{arrows}

\usepackage{color}
\usepackage{url}

\usepackage[colorlinks]{hyperref}
\hypersetup{%
        plainpages=true,
        breaklinks=true,
        hypertexnames=false,
        pageanchor=true,
        colorlinks=true,
        linkcolor={blue},
        citecolor={magenta},
        urlcolor={blue},
        anchorcolor={black}
      }

\usepackage{mleftright} 

\newcommand{\figref}[1]{\mbox{Fig.~\ref{#1}}}
\newcommand{\tabref}[1]{\mbox{Table~\ref{#1}}}

\renewcommand{\eqref}[1]{\mbox{Eq.~(\ref{#1})}}

\newcommand{\ket}[1]{|#1\rangle}

\newcommand{\abs}[1]{\left|#1\right|}

\newcommand{\be}{\begin{equation}}
\newcommand{\ee}{\end{equation}}
\newcommand{\bea}{\begin{eqnarray}}
\newcommand{\eea}{\end{eqnarray}}

\begin{document}

\title{Large collective Lamb shift of two distant superconducting artificial atoms}

\author{P.~Y.~Wen}
\thanks{These authors contributed equally}
\affiliation{Department of Physics, National Tsing Hua University, Hsinchu 30013, Taiwan}
\affiliation{Center for Quantum Technology, National Tsing Hua University, Hsinchu 30013, Taiwan}

\author{K.-T.~Lin}
\thanks{These authors contributed equally}
\affiliation{CQSE, Department of Physics, National Taiwan University, Taipei 10617, Taiwan}

\author{A.~F.~Kockum}
\thanks{These authors contributed equally}
\affiliation{Department of Microtechnology and Nanoscience, Chalmers University of Technology, 412 96 Gothenburg, Sweden}
\affiliation{Theoretical Quantum Physics Laboratory, RIKEN Cluster for Pioneering Research, Wako-shi, Saitama 351-0198, Japan}

\author{B.~Suri}
\affiliation{Department of Instrumentation and Applied Physics, Indian Institute of Science, Bengaluru 560012, India}
\affiliation{Department of Microtechnology and Nanoscience, Chalmers University of Technology, 412 96 Gothenburg, Sweden}

\author{H.~Ian}
\affiliation{Institute of Applied Physics and Materials Engineering, University of Macau, Macau}
\affiliation{UMacau Zhuhai Research Institute, Zhuhai, Guangdong, China}

\author{J.~C.~Chen}
\affiliation{Department of Physics, National Tsing Hua University, Hsinchu 30013, Taiwan}
\affiliation{Center for Quantum Technology, National Tsing Hua University, Hsinchu 30013, Taiwan}

\author{S.~Y.~Mao}
\affiliation{Institute of Electro-Optical Engineering, National Chiao Tung University, Hsinchu 30013, Taiwan}

\author{C.~C.~Chiu}
\affiliation{Department of Electrical Engineering, National Tsing Hua University, Hsinchu 30013, Taiwan}

\author{P.~Delsing}
\affiliation{Department of Microtechnology and Nanoscience, Chalmers University of Technology, 412 96 Gothenburg, Sweden}

\author{F.~Nori}
\affiliation{Theoretical Quantum Physics Laboratory, RIKEN Cluster for Pioneering Research, Wako-shi, Saitama 351-0198, Japan}
\affiliation{Physics Department, The University of Michigan, Ann Arbor, Michigan 48109-1040, USA}

\author{G.-D.~Lin}
\affiliation{CQSE, Department of Physics, National Taiwan University, Taipei 10617, Taiwan}

\author{I.-C.~Hoi}
\email[e-mail:]{ichoi@phys.nthu.edu.tw}
\affiliation{Department of Physics, National Tsing Hua University, Hsinchu 30013, Taiwan}
\affiliation{Center for Quantum Technology, National Tsing Hua University, Hsinchu 30013, Taiwan}

\date{\today}

\begin{abstract}

Virtual photons can mediate interaction between atoms, resulting in an energy shift known as a collective Lamb shift. Observing the collective Lamb shift is challenging, since it can be obscured by radiative decay and direct atom-atom interactions. Here, we place two superconducting qubits in a transmission line terminated by a mirror, which suppresses decay. We measure a collective Lamb shift reaching 0.8\% of the qubit transition frequency and exceeding the transition linewidth. We also show that the qubits can interact via the transmission line even if one of them does not decay into it.


\end{abstract}

\maketitle


\paragraph*{Introduction.}

In 1947, when attempting to pinpoint the fine structure of the hydrogen atom, Lamb and Retherford~\cite{Lamb1947} discovered a small energy difference between the levels $2 S_{1/2}$ and $2 P_{1/2}$, which were thought to be degenerate according to Dirac's theory of electrons. This energy difference between the two levels can be understood when vacuum fluctuations are included in the picture, as was verified later by self-energy calculations in the framework of quantum field theory~\cite{Bethe1947, Welton1948, Cohen-Tannoudji1986}. Briefly put, a hydrogen atom will emit photons which are instantaneously reabsorbed; while these ``virtual'' photons are not detectable by themselves, they leave their traces in the Lamb shift.

The hydrogen atoms that Lamb and Retherford used for their experiment were obtained from molecular hydrogen through tungsten catalyzation. Since the conversion rate for this process was very low, the $2 S_{1/2}$ level was only populated in a few atoms. Hence, the observable effects of virtual photon processes were limited to self-interaction; exchanges of virtual photons between atoms could not be detected. However, it was later realized that atom-atom interaction mediated by virtual photons also gives rise to an energy shift, referred to as a collective, or cooperative, Lamb shift~\cite{Fain1959, Lehmberg1970, Arecchi1970, Friedberg1973, Scully2010}. The atom-atom interaction also underpins the collective decay known as Dicke superradiance~\cite{Dicke1954, Shammah2018}.

There are several obstacles impeding the experimental observation of the collective Lamb shift. The shift can be enhanced by using many atoms, but, if these atoms are too close together, direct atom-atom interactions (not via virtual photons) can obscure the effect. Furthermore, the interaction giving rise to the collective Lamb shift is relatively weak in three dimensions, and the shift can also be hidden by the radiative linewidth (e.g., due to the collective decay). Despite these obstacles, there have been a few experimental demonstrations of collective Lamb shifts: in xenon gas~\cite{Garrett1990}, iron nuclei~\cite{Rohlsberger2010}, rubidium vapor~\cite{Keaveney2012}, strontium ions~\cite{Meir2014}, cold rubidium atoms~\cite{Roof2016}, and potassium vapor~\cite{Peyrot2018}. Mostly, these experiments used developments in atomic trapping and cooling~\cite{Lukin2003} that have enabled higher densities of atomic ensembles, leading to a strong coupling between atomic condensates and cavity fields~\cite{Brennecke2007, Colombe2007}. An improved theoretical understanding~\cite{Scully2006, Svidzinsky2008, Scully2009} of collective Dicke states also aided some of the experiments.

With the single exception of Ref.~\cite{Meir2014}, these previous experiments all required a large number of atoms to demonstrate a collective Lamb shift. The experiment of Ref.~\cite{Meir2014} only used two atoms, but the measured shift was small, 0.2\% of the transition linewidth. In this Letter, we demonstrate a large collective Lamb shift for two artificial atoms that significantly exceeds the linewidth and reaches 0.8\% of the atomic transition frequency. 

\begin{figure}
\includegraphics[width=\linewidth]{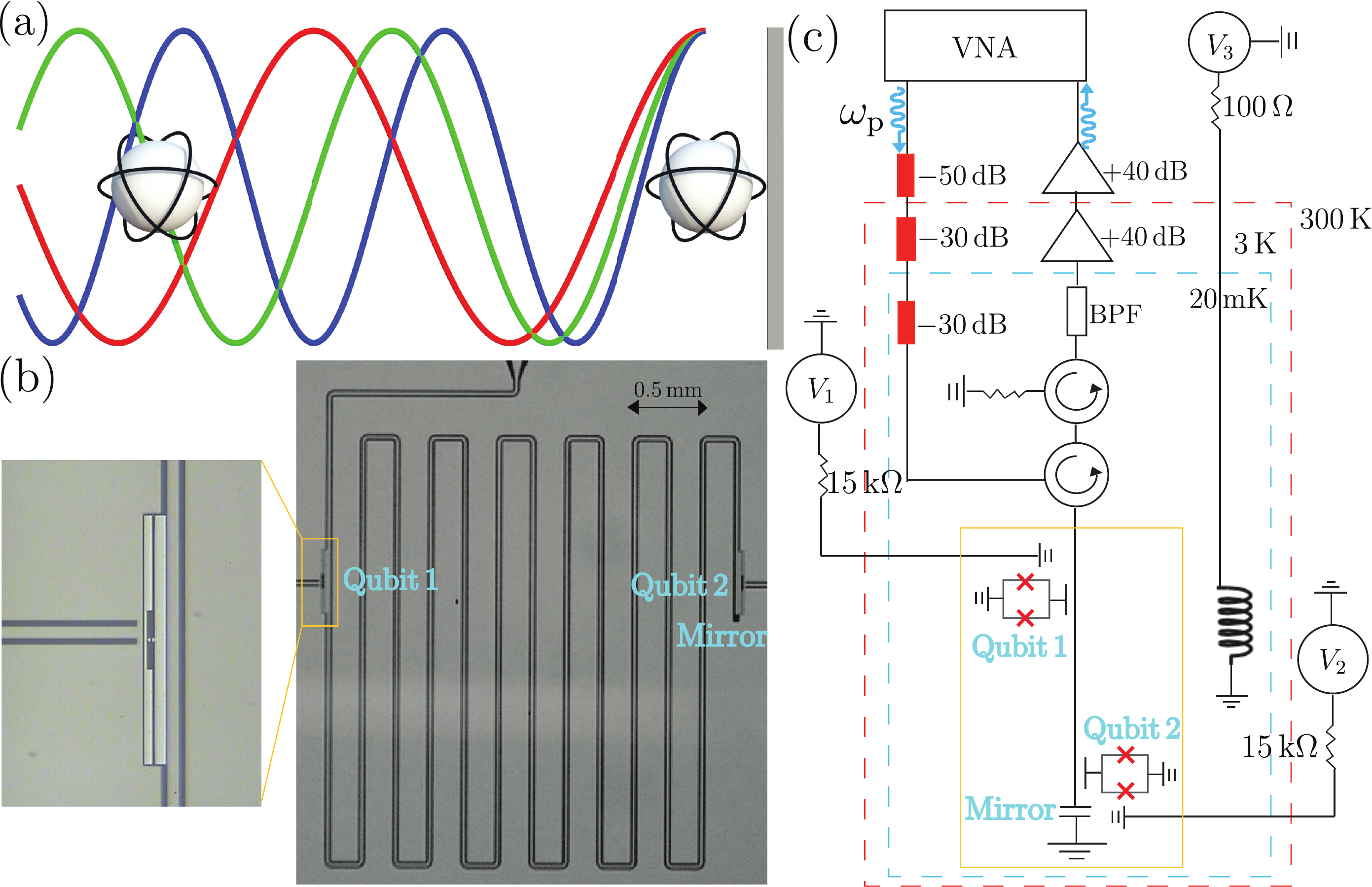}
\caption{The experimental setup. 
(a) A conceptual sketch of the setup. Two atoms are placed in front of a mirror and interact via virtual photons of different frequencies.
(b) A photo of the device. Qubit 1 (shown in the zoom-in on the left; the two long bright parts form the qubit capacitance and the gap in the middle between them is bridged by two Josephson junctions forming a SQUID loop) is placed $L\simeq\unit[33]{mm}$ from Qubit 2, which sits at the end of the transmission line (i.e., at the mirror). The characteristic impedance of the transmission line is $Z_0 \simeq \unit[50]{\Omega}$. The relatively long distance $L$ makes it easier to tune Qubit 1 between nodes and anti-nodes of the electromagnetic (EM) field in the transmission line by tuning the qubit transition frequency. This tuning can be used to calibrate the velocity $v$ of the EM field in the transmission line~\cite{SupMat}.
(c) A sketch of the signal routing for the experiment. Each qubit frequency can be tuned by local magnetic fields via local voltage biases ($V_1$, $V_2$) and both frequencies can be tuned by a global magnetic field from a superconducting coil controlled by $V_3$. For measurements, a coherent signal at frequency $\omega_{\rm p}$ is generated by a vector network analyzer (VNA) at room temperature and fed through attenuators (red squares) to the sample, which sits in a cryostat cooled to $\unit[20]{mK}$ to avoid thermal fluctuations affecting the experiment. The reflected signal passes a bandpass filter (BPF) and amplifiers, and is then measured with the VNA.
\label{fig:Device}}
\end{figure}

Our experimental setup, depicted in \figref{fig:Device}, is a superconducting quantum circuit~\cite{You2011, Gu2017} with two transmon qubits~\cite{Koch2007} coupled to a one-dimensional (1D) waveguide. In such superconducting circuits, strong~\cite{Wallraff2004, Astafiev2010, Gu2017}, and even ultrastrong~\cite{Niemczyk2010, Forn-Diaz2017, Kockum2019, Forn-Diaz2019}, coupling can be engineered between the qubits and photons in the waveguide. Compared to three-dimensional setups, the 1D version strengthens the interaction between qubits and reduces the decay into unwanted modes. These features have enabled many important quantum-optical experiments in 1D waveguide QED in superconducting circuits in the past decade~\cite{Gu2017, Roy2017, Astafiev2010, Hoi2011, Hoi2012, VanLoo2013, Hoi2013a, Hoi2015, Forn-Diaz2017, Liu2017, Wen2018, Mirhosseini2018, Sundaresan2019, Mirhosseini2019} and inspired a wealth of theoretical studies for this platform~\cite{Gu2017, Roy2017, Shen2005, Chang2007, Zhou2008, Dong2009, Zheng2010, Chen2011, Gonzalez-Tudela2011, Chang2012, Koshino2012, Stannigel2012, Fan2013, Lalumiere2013, Peropadre2013, Tufarelli2013, Laakso2014, Lindkvist2014, Sathyamoorthy2014, Shahmoon2014, Fang2015, Paulisch2016, Pichler2016, Gonzalez-Tudela2017, Guo2017, Kockum2018}.

As shown in \figref{fig:Device}, the transmission line to which the qubits couple is terminated in a capacitive coupling to ground, which is equivalent to placing a mirror in a waveguide. The presence of this mirror separates our experiment from that of Ref.~\cite{VanLoo2013}, where two superconducting qubits were coupled to an open transmission line. In such an open waveguide, the connection between collective decay and the collective Lamb shift entails that the shift always will be smaller than the linewidth~\cite{Lalumiere2013}, and the measurements of elastic scattering in Ref.~\cite{VanLoo2013} could thus not resolve the collective Lamb shift. Although a splitting in the fluorescence spectrum (inelastic scattering) indicated the presence of the collective Lamb shift, it is not straightforward to extract the size of the shift from the size of the splitting~\cite{VanLoo2013, Lalumiere2013}. In our setup, the presence of the mirror introduces interference effects that suppresses the collective decay more than the collective Lamb shift~\cite{SupMat, Kockum2018}, allowing us to clearly resolve the shift in simple reflection measurements of elastic scattering. Interestingly, it turns out that these interference effects allow us to couple the two qubits via the transmission line even when one of the qubits is unable to relax into the transmission line.

\paragraph*{Device and characterization.}

\begin{figure}
\includegraphics[width=\linewidth]{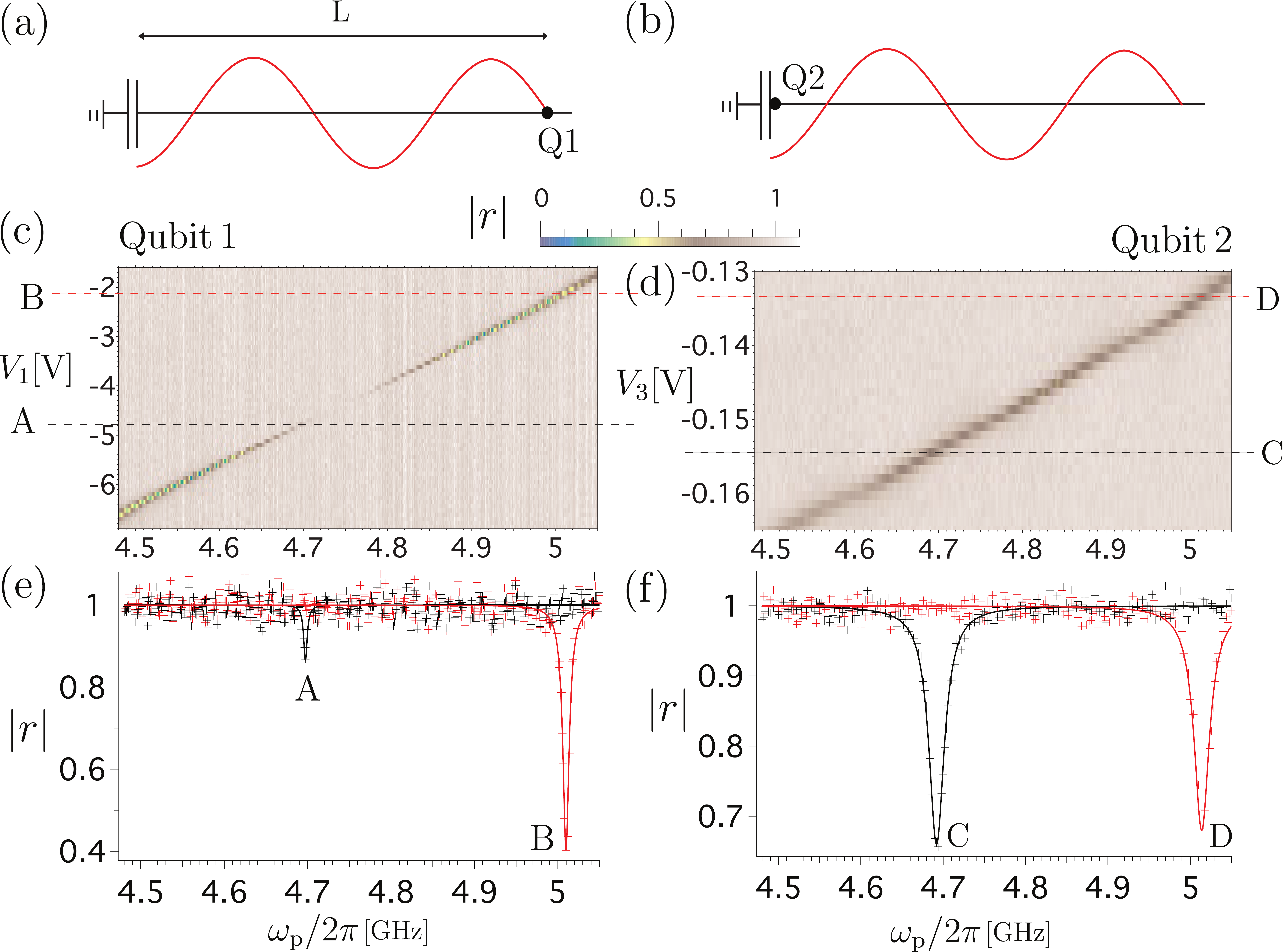}
\caption{Single-tone spectroscopy of the individual qubits. 
(a), (b) Electromagnetic mode structure (red curve) in the transmission line seen by Qubit 1 (Q1) and Qubit 2 (Q2), respectively.
(c), (d) Amplitude reflection coefficient $\abs{r}$ for a weak coherent probe as a function of probe frequency $\omega_{\rm p}$ and qubit transition frequency (controlled by the voltages $V_1$ and $V_3$ for Qubit 1 and Qubit 2, respectively). For Qubit 1, panel (c) shows how the response disappears when the qubit ends up at a node for the EM field around $\unit[4.75]{GHz}$. During these measurements, the frequency of the other qubit is tuned far from resonance with the probe.
(e), (f) Linecuts from panels (c) and (d) as indicated. Crosses are experimental data and solid curves are fits following Ref.~\cite{Hoi2013a}. The extracted parameters are given in Table S1 in the supplementary material~\cite{SupMat}. The linewidth of the dip, which occurs at the resonance $\omega_{\rm p} = \omega_{10}$, is set by the qubit decoherence rate $\gamma = \Gamma / 2 + \gamma_\phi$, where $\Gamma$ is the relaxation rate and $\gamma_\phi$ is the pure dephasing rate. Relaxation into other channels than the transmission line will affect the extracted value of $\gamma_\phi$. The depth of the dip is set by the ratio $\Gamma / \gamma_\phi$; since $\Gamma$ decreases close to the node of the field, the dip in linecut A is more shallow than that in B.
\label{fig:FluxDependence}}
\end{figure}

In our device, the interqubit separation $L$ is fixed. However, we can vary the qubit transition frequencies $\omega_{10}$ by applying a local magnetic flux [see \figref{fig:Device}(c)] and thus change the effective distance $L / \lambda$, where the wavelength $\lambda = 2 \pi v / \omega_{10}$ and $v$ is the propagation velocity of the electromagnetic (EM) field in the waveguide~\cite{Hoi2015}. Since Qubit 2 is placed next to the mirror, it will always be at an antinode of the voltage field in the waveguide [see \figref{fig:FluxDependence}(b)]. Qubit 1, on the other hand, can be tuned to a voltage node. In this case, Qubit 1 will not couple to the waveguide at its transition frequency, and thus will not contribute to any decay~\cite{Hoi2015, Kockum2018}. However, the collective Lamb shift arises due to emission and absorption of virtual photons in all other modes of the continuum in the waveguide, which results in an interaction of strength $\Delta$ between the qubits. This interaction (collective Lamb shift) leads to an avoided level crossing between the two qubits, which shows up as a frequency splitting of $2\Delta$ in reflection measurements of the system using a weak coherent probe at frequency $\omega_{\rm p}$. Our experiment thus clearly demonstrates how the collective Lamb shift has contributions from virtual photons of many frequencies.

We first characterize each of the two transmon qubits through spectroscopy. We detune the transition frequency of one of the qubits far away and measure the amplitude reflection coefficient $\abs{r}$ of a weak coherent probe tone (i.e., the probe Rabi frequency $\Omega_{\rm p}$ is much smaller than the decoherence rate $\gamma$ of the qubit) as a function of the flux controlling the other qubit's transition frequency and of the probe frequency $\omega_{\rm p}$. The results are shown in \figref{fig:FluxDependence} (Qubit 1 in the left column and Qubit 2 in the right column). For Qubit 1, which is placed at a distance $L$ from the mirror, the spectroscopy data in \figref{fig:FluxDependence}(c) shows a linewidth narrowing [compare the linecuts A and B from \figref{fig:FluxDependence}(c), plotted in \figref{fig:FluxDependence}(e)] and a disappearing response around $\unit[4.75]{GHz}$. At this frequency, the effective distance between Qubit 1 and the mirror is $L = 7 \lambda / 4$, which places the qubit at a node for the EM field in the transmission line, as illustrated in \figref{fig:FluxDependence}(a), and thus effectively decouples the qubit from the transmission line, reducing its relaxation rate to zero~\cite{Hoi2015}. Qubit 2, on the other hand, is always at an antinode of the EM field in the transmission line [\figref{fig:FluxDependence}(b)] and thus has an equally strong response at all frequencies [\figref{fig:FluxDependence}(d), (f)].

We perform further spectroscopy in the full range $\unit[4-8]{GHz}$, which is the bandwidth of the cryogenic low-noise amplifier in our experimental setup. The maximum qubit frequency is outside this bandwidth. This data is presented in the supplementary material~\cite{SupMat}. From these measurements, we extract~\cite{Hoi2013a} the qubit relaxation rate $\Gamma$ into the transmission line, the pure dephasing rate $\gamma_\phi$ (which also contains contributions from relaxation to other channels), and the speed of light in the transmission line. We further use two-tone spectroscopy, driving at the qubit frequency $\omega_{10}$ and probing around the transition frequency $\omega_{21}$ from the first excited state to the second excited state, to determine the anharmonicity of the qubits. All extracted and derived parameters are summarized in \tabref{tab:Parameters}.

\begin{table*}
\centering
\begin{tabular}{| c |c | c | c | c | c | c | c | }
\hline
Qubit & $E_C / h$ [GHz] & $\omega_{10} / 2\pi$ [GHz]  & $\Gamma / 2\pi$ [MHz] & $\gamma_\phi / 2\pi$ [MHz] & $\gamma / 2\pi$ [MHz] & $\beta$ & $v$ [$10^8$m/s] \\
\hline
$Q1$ & $0.324$ & (Antinode) $4.068$ & $27.18$ & $2.15$ & $15.74$ & $0.717$ & $0.8948$ \\
$Q2$ & $0.406$ & (Antinode) $4.746$ & $28.03$ & $2.785$ & $16.8$ & $0.696$ & \\
\hline
\end{tabular}
\caption{Extracted and derived qubit parameters. We extract $\omega_{10}$, $\Gamma$ and $\gamma$ from fitting the spectroscopic magnitude and phase data according to Ref.~\cite{Hoi2013a}. Note that the effective relaxation rate $\Gamma$ at an antinode is twice what the relaxation rate would be in an open transmission line. The velocity $v$ is extracted by finding multiple nodes of the field for Qubit 1~\cite{SupMat}. From the two-tone spectroscopy, we extract the anharmonicity, which approximately equals the charging energy $E_C$ of the transmon qubits. We calculate $\gamma_\phi$ from $\Gamma$ and $\gamma$, and the ratio $\beta = C_{\rm c} / C_\Sigma$ between the coupling capacitance $C_{\rm c}$ to the transmission line and the qubit capacitance $C_\Sigma$ from $\Gamma$ and $E_C$.
\label{tab:Parameters}}
\end{table*} 

\paragraph*{Collective Lamb shift.}

\begin{figure*}
\includegraphics[width=\linewidth]{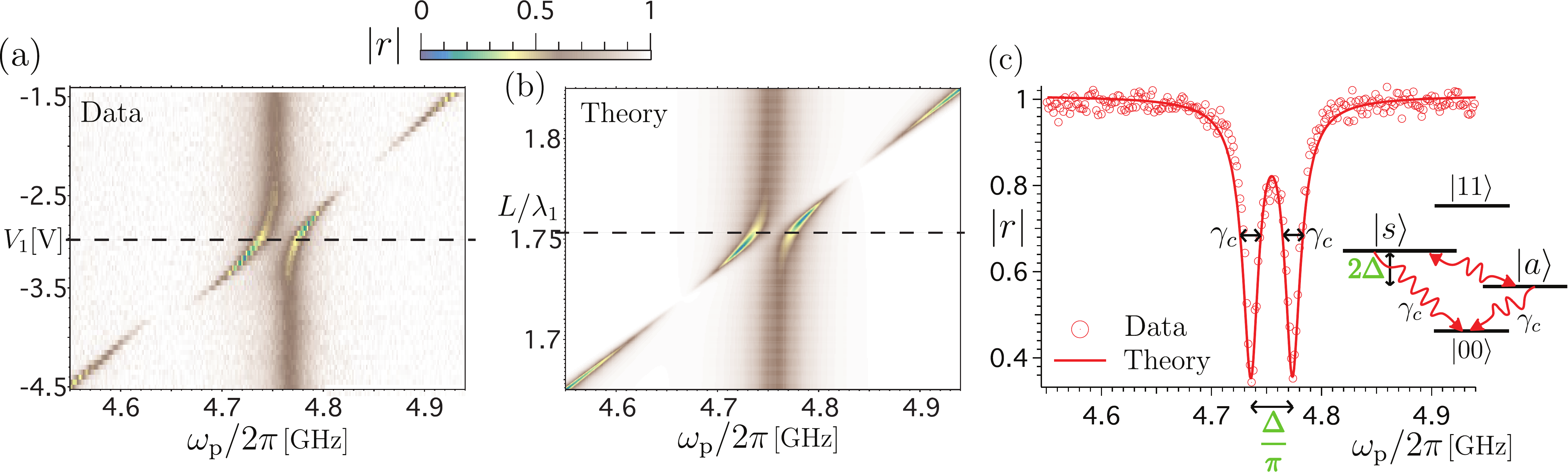}
\caption{Collective Lamb shift.
(a) The amplitude reflection coefficient for a weak probe as a function of the probe frequency $\omega_{\rm p}$ and the transition frequency of Qubit 1 (controlled through the voltage $V_1$). The frequency of Qubit 2 is fixed at $\omega_{10} = \unit[4.75]{GHz}$ and the frequency of Qubit 1 is tuned through resonance with this frequency.
(b) Theory simulation~\cite{SupMat} of the single-tone spectroscopy data in panel (a). The simulation is done with previously fitted parameters from \tabref{tab:Parameters}, with the exceptions of the free paramaters $\beta_1 = 0.81$, $\beta_2 = 0.766$, and $\gamma_\phi  / 2 \pi = \unit[2.3]{MHz}$ for Qubit 1, which all are close to the values in \tabref{tab:Parameters}. The agreement between the data in (a) and the simulation in (b) is excellent.
(c) A linecut of the data and theory, marked by the dashed line in panels (a) and (b), at the point where the two qubits are on resonance and the collective Lamb shift $2 \Delta$ is most clearly visible. From this figure, we extract a collective Lamb shift of $2 \Delta \simeq 2 \pi \times \unit[38]{MHz}$. The inset on the right shows the level structure of the eigenstates of the qubits that are coupled through the collective Lamb shift. The two dips in the reflection corresponds to the symmetric and anti-symmetric eigenstates $\ket{s}$ and $\ket{a}$. Each of these two states have the same decay rate, giving a linewidth $\gamma_c$ smaller than the collective Lamb shift due to the presence of the mirror, as explained in the text.
\label{fig:VacuumRabi}}
\end{figure*}

We now turn to experiments where both qubits are involved and the collective Lamb shift is measured. We fix the transition frequency of Qubit 2 to $\omega_{10} / 2 \pi = \unit[4.75]{GHz}$, the frequency at which Qubit 1 is at a node of the EM field [see \figref{fig:FluxDependence}(c)]. We then tune the frequency of Qubit 1 to values around this point and measure the reflection of a weak probe signal on the system for frequencies close to $\omega_{10}$. The results of these measurements are displayed in \figref{fig:VacuumRabi}(a). We observe a clear anti-crossing between the vertical resonance, corresponding to Qubit 2, and the diagonal resonance, corresponding to Qubit 1. The observation of this anti-crossing indicates that the two qubits are coupled on resonance with strength $\Delta$ through a coherent interaction, which must be mediated by the transmission line since the qubits are distant from each other. The minimum size of the separation, shown in the linecut in \figref{fig:VacuumRabi}(c), is $2 \Delta \simeq 2 \pi \times \unit[38]{MHz}$.

If the qubits were uncoupled, they would have eigenstates $\ket{00}$, $\ket{01}$, $\ket{10}$, and $\ket{11}$, with energies $0$, $\hbar \omega_{10}$, $\hbar \omega_{10}$, and $2 \hbar \omega_{10}$, respectively. Here, $0$ and $1$ denote ground and excited states of a single qubit, respectively; the first number in the kets is for Qubit 1 and the second number is for Qubit 2. Due to the coupling, the eigenstates $\ket{01}$ and $\ket{10}$ are replaced by the symmetric and anti-symmetric eigenstates $\ket{s} = \frac{1}{\sqrt{2}} \mleft( \ket{01} + \ket{10} \mright)$ and $\ket{a} = \frac{1}{\sqrt{2}} \mleft( \ket{01} - \ket{10} \mright)$, respectively, with eigenenergies $\hbar \mleft(\omega_{10} \pm \Delta \mright)$~\cite{Lalumiere2013}. When the coupling is due to virtual photons, as in our experiment, this thus gives a collective Lamb shift of $2 \hbar \Delta$, as illustrated in the inset in \figref{fig:VacuumRabi}(c).

If the two qubits were placed in an open transmission line, it would not be possible to observe the collective Lamb shift in this measurement, since each of the two resonances would have a linewidth set by a relaxation rate $\Gamma = 2 \Delta$~\cite{Lalumiere2013}. This is not easily circumvented, since it is the coupling to the transmission line of the two qubits that determines both the relaxation into the transmission line and the strength of the interaction that is mediated via the transmission line. However, the presence of the mirror in our setup breaks this close connection between the linewidth and the collective Lamb shift. In our setup, the collective Lamb shift is given by~\cite{Kockum2018, SupMat}
\be
2 \Delta = \Gamma_0 \mleft\{ \sin \mleft[ \frac{\omega_{10}}{v}(x_1 + x_2) \mright] + \sin \mleft[ \frac{\omega_{10}}{v} \abs{x_1 - x_2} \mright] \mright\},
\ee
where $x_j$ denotes the distance of Qubit $j$ from the mirror and $\Gamma_0 = \sqrt{\Gamma_1 (\omega_{10}) \Gamma_2 (\omega_{10})}$, with $\Gamma_j (\omega)$ the bare relaxation rate of Qubit $j$ at frequency $\omega$ into an open transmission line. This is calculated using the standard master-equation approach with the Born-Markov approximation and tracing out the photonic modes of the transmission line~\cite{SupMat, Carmichael1999}. When $x_2 = 0$ and $x_1$ corresponds to Qubit 1 being at a node of the field in the transmission line, as in \figref{fig:VacuumRabi}, the collective Lamb shift becomes $2 \Delta = 2 \Gamma_0$. However, since Qubit 1 is at a node, both the effective relaxation rate of Qubit 1 and the collective decay rate of the two qubits becomes zero. The only contribution from relaxation to the linewidths for the states $\ket{s}$ and $\ket{a}$ is half of $2 \Gamma_2$, the effective relaxation rate of Qubit 2. In this experiment, we used $\Gamma_1 \approx \Gamma_2$ (giving a shift $2 \Delta \approx 2 \Gamma_2$ and a linewidth $\gamma_c \approx \Gamma_2$), but we note that the collective Lamb shift could be made many times larger than the linewidths by instead designing the qubits such that $\Gamma_1 \gg \Gamma_2$.

The fact that we can measure the collective Lamb shift even though Qubit 1 ostensibly is decoupled from the transmission line confirms several predictions about how virtual photons influence relaxation and qubit-qubit interaction. The relaxation from Qubit 1 is stimulated by virtual photons in the transmission line at the transition frequency $\omega_{10}$. The relaxation is suppressed when Qubit 1 is placed at a node for the virtual photons at this frequency~\cite{Hoi2015}. However, Qubit 1 is clearly coupled via virtual photons to Qubit 2. Thus, the virtual photons mediating this coupling, and causing the collective Lamb shift, must have frequencies that are not equal to $\omega_{10}$. In fact, the coupling is given by a sum over all virtual modes at frequencies separate from $\omega_{10}$~\cite{Lalumiere2013}.

\begin{figure}
\includegraphics[width=\linewidth]{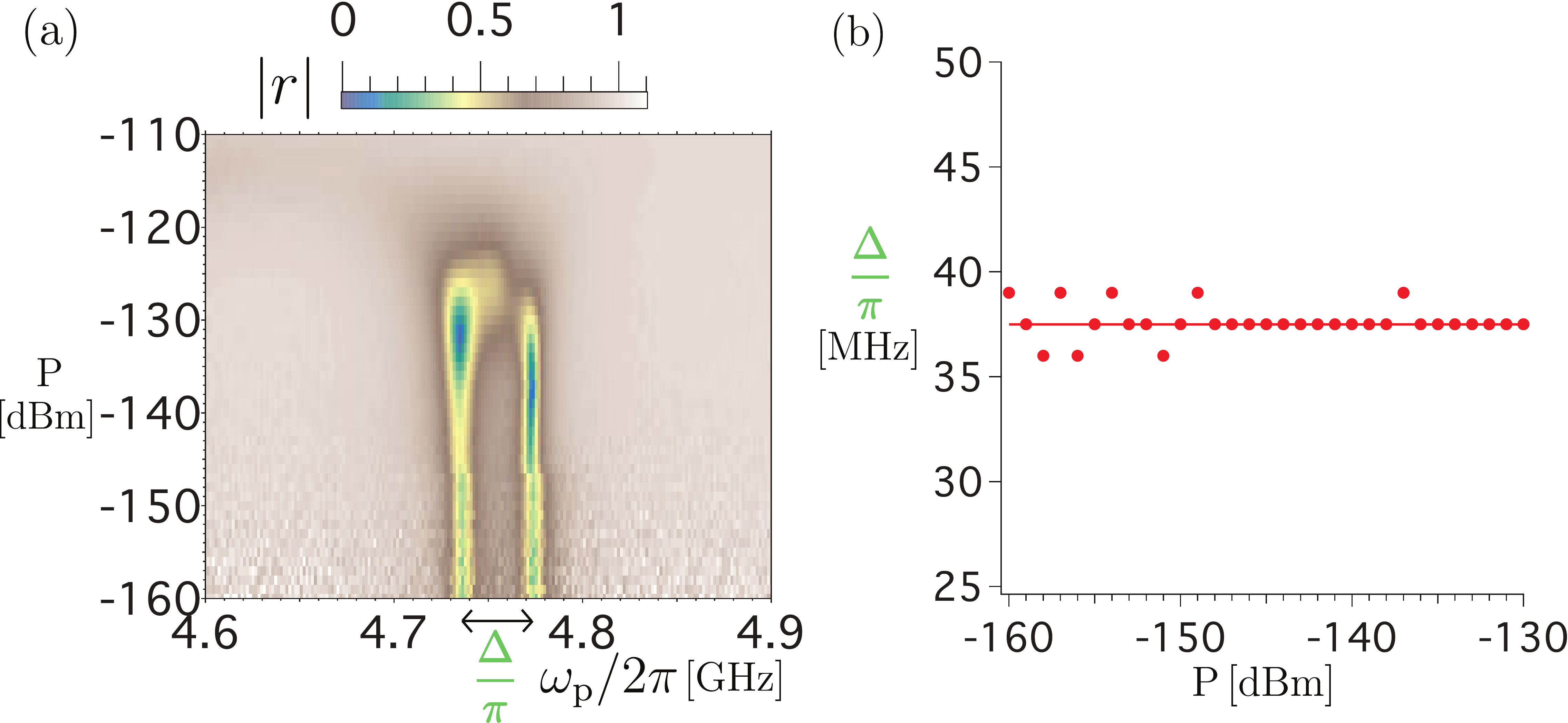}
\caption{Energy shift as a function of input power.
(a) Amplitude reflection coefficient $\abs{r}$ as a function of probe frequency $\omega_{\rm p}$ and probe power $P$ for both qubit frequencies fixed at $\omega_{10} = \unit[4.75]{GHz}$. The data agrees very well with theoretical simulations~\cite{SupMat}. At high probe powers, the qubits are saturated and most photons are simply reflected from the mirror, resulting in $\abs{r} \approx 1$.
(b) The extracted splitting $2 \Delta$ from panel (a) as a function of $P$. The splitting is clearly independent of the input power in this wide range.
\label{fig:PowerDependence}}
\end{figure}

Finally, we note that there are several processes, with real photons, where a strong drive shifts or dresses energy levels of qubits to create an effect that could look similar to what we have observed. To rule out such effects, e.g., the Mollow triplet~\cite{Mollow1969} and Autler-Townes splitting~\cite{Autler1955}, we measure $\Delta$ as a function of the power $P$ of the coherent probe. The results are shown in \figref{fig:PowerDependence}. Clearly, the energy shift $\Delta$ is independent of $P$ (before the power is high enough to saturate the qubits), indicating that the collective Lamb shift we measure really is due to virtual photons.

\paragraph*{Summary and outlook.}

In this Letter, we demonstrated a large collective Lamb shift with two distant superconducting qubits in front of an effective mirror in a 1D transmission-line waveguide. Using interference effects due to the mirror, we overcame previous limitations on the size of the shift compared to the linewidth, allowing us to observe a shift reaching 0.8\% of the qubit transition frequency and exceeding the transition linewidth. We explained how future experiments could increase the shift relative to the linewidth even more. This experiment also demonstrated that a qubit can couple to another qubit via the transmission line even though the first qubit is prevented from decaying into the transmission line. These results give further insight into how virtual photons affect both atomic relaxation rates and inter-atomic coupling, and how these effects can be controlled using interference, which could have applications for designing, e.g., devices that process quantum information.


\begin{acknowledgments}

\paragraph*{Acknowledgements.}

I.-C.H. and J.C.C. would like to thank I. A. Yu and C.-Y. Mou for fruitful discussions.
This work was financially supported by the Center for Quantum Technology from the Featured Areas Research Center Program within the framework of the Higher Education Sprout Project by the Ministry of Education (MOE) in Taiwan. 
I.-C.H.~acknowledges financial support from the MOST of Taiwan under project 107-2112-M-007-008-MY3.
B.S., A.F.K., and P.D. acknowledge support from the Knut and Alice Wallenberg Foundation.
G.-D.L acknowledges support from the MOST of Taiwan under Grant No.~105-2112-M-002-015-MY3 and National 
Taiwan University under Grant No.~NTUCC-108L893206.
H. I. acknowledges the support by FDCT of Macau under grant 065/2016/A2, by University of Macau under grant MYRG2018-00088-IAPME, and by NNSFC under grant 11404415.
J.C.C. acknowledges financial support from the MOST of Taiwan under project 107-2112-M-007-003-MY3.
F.N.~acknowledges support from the MURI Center for Dynamic Magneto-Optics via the Air Force Office of Scientific Research (AFOSR) award No.~FA9550-14-1-0040, the Army Research Office (ARO) under grant No.~W911NF-18-1-0358, the Asian Office of Aerospace Research and Development (AOARD) grant No.~FA2386-18-1-4045, the Japan Science and Technology Agency (JST) through the Q-LEAP program, the ImPACT program, and CREST Grant No.~JPMJCR1676, the Japan Society for the Promotion of Science (JSPS) through the JSPS-RFBR grant No.~17-52-50023 and the JSPS-FWO grant No.~VS.059.18N, the RIKEN-AIST Challenge Research Fund, and the John Templeton Foundation.

\end{acknowledgments}

\bibliography{LambShiftRefs}

\end{document}


\title{Supplementary Material for ``Large collective Lamb shift of two distant superconducting artificial atoms''}

\author{P.~Y.~Wen}
\thanks{These authors contributed equally}
\affiliation{Department of Physics, National Tsing Hua University, Hsinchu 30013, Taiwan}
\affiliation{Center for Quantum Technology, National Tsing Hua University, Hsinchu 30013, Taiwan}

\author{K.-T.~Lin}
\thanks{These authors contributed equally}
\affiliation{Department of Physics, National Taiwan University, Taipei 10617, Taiwan}

\author{A.~F.~Kockum}
\thanks{These authors contributed equally}
\affiliation{Department of Microtechnology and Nanoscience, Chalmers University of Technology, 412 96 Gothenburg, Sweden}
\affiliation{Theoretical Quantum Physics Laboratory, RIKEN Cluster for Pioneering Research, Wako-shi, Saitama 351-0198, Japan}

\author{B.~Suri}
\affiliation{Department of Instrumentation and Applied Physics, Indian Institute of Science, Bengaluru 560012, India}
\affiliation{Department of Microtechnology and Nanoscience, Chalmers University of Technology, 412 96 Gothenburg, Sweden}

\author{H.~Ian}
\affiliation{Institute of Applied Physics and Materials Engineering, University of Macau, Macau}
\affiliation{UMacau Zhuhai Research Institute, Zhuhai, Guangdong, China}

\author{J.~C.~Chen}
\affiliation{Department of Physics, National Tsing Hua University, Hsinchu 30013, Taiwan}
\affiliation{Center for Quantum Technology, National Tsing Hua University, Hsinchu 30013, Taiwan}

\author{S.~Y.~Mao}
\affiliation{Institute of Electro-Optical Engineering, National Chiao Tung University, Hsinchu 30013, Taiwan}

\author{C.~C.~Chiu}
\affiliation{Department of Electrical Engineering, National Tsing Hua University, Hsinchu 30013, Taiwan}

\author{P.~Delsing}
\affiliation{Department of Microtechnology and Nanoscience, Chalmers University of Technology, 412 96 Gothenburg, Sweden}

\author{F.~Nori}
\affiliation{Theoretical Quantum Physics Laboratory, RIKEN Cluster for Pioneering Research, Wako-shi, Saitama 351-0198, Japan}
\affiliation{Physics Department, The University of Michigan, Ann Arbor, Michigan 48109-1040, USA}

\author{G.-D.~Lin}
\affiliation{Department of Physics, National Taiwan University, Taipei 10617, Taiwan}

\author{I.-C.~Hoi}
\email[e-mail:]{ichoi@phys.nthu.edu.tw}
\affiliation{Department of Physics, National Tsing Hua University, Hsinchu 30013, Taiwan}
\affiliation{Center for Quantum Technology, National Tsing Hua University, Hsinchu 30013, Taiwan}

\date{\today}

\maketitle

\tableofcontents

\renewcommand{\thefigure}{S\arabic{figure}}
\renewcommand{\thesection}{S\arabic{section}}
\renewcommand{\theequation}{S\arabic{equation}}
\renewcommand{\thetable}{S\arabic{table}}
\renewcommand{\bibnumfmt}[1]{[S#1]}
\renewcommand{\citenumfont}[1]{S#1}


\section{Derivation of master equation and qubit-qubit interaction}

In this section, we outline the derivation of qubit-qubit coupling through virtual photons in the continuum of photonic modes in a 1D transmission line terminated by a mirror. We consider $N$ transmon qubits, placed at positions $x_n$ in the transmission line. The coordinate $x_n$ measures the distance from qubit $n$ to the mirror at $x = 0$. The Hamiltonian for this system can be expressed as $H = H_{\rm S} + H_{\rm B} + H_{\rm int}$ with
\bea
H_{\rm S} &=& \sum_{n = 1}^N \hbar \omega_{10}^n \sigma_n^ + \sigma_n^-, \\
H_{\rm B} &=& \int d \omega \hbar \omega a_\omega^\dag a_\omega, \\
H_{\rm int} &=& i \sum_{n = 1}^N \int d \omega \hbar g_n (\omega) \cos \mleft( k_\omega x_n \mright) \mleft( a_\omega \sigma_n^+ - \sigma_n^- a_\omega^\dag \mright).
\eea
Here, $H_{\rm S}$ is the bare Hamiltonian of the qubits, with $\sigma_n^+$ ($\sigma_n^-$) the raising (lowering) operator of qubit $n$ and $\omega_{10}^{n}$ the transition frequency of qubit $n$. The bare Hamiltonian for the continuum of photonic modes in the transmission line is given by $H_{\rm B}$, where $a_\omega^\dag$ ($a_\omega$) is the creation (annihilation) operator for excitations at mode frequency $\omega$. The interaction between the qubits and the photons is described by $H_{\rm int}$, where the interaction strength is given by~\cite{Hoi2015}
\be
g_n (\omega) = e \beta_n \mleft( \frac{E_{\rm J}^{(n)}}{8 E_{\rm C}^{(n)}} \mright)^{1/4} \sqrt{\frac{2 Z_0 \omega}{\hbar \pi}},
\ee
where $\beta_n = C_{\rm c}^n /C_\Sigma^n$ is the ratio between the coupling capacitance $C_{\rm c}^n$ to the transmission line and the qubit capacitance $C_\Sigma^n$ for qubit $n$, $E_{\rm C}^{(n)}$ and $E_{\rm J}^{(n)}$ are the charging and Josephson energies, respectively, of qubit $n$, $e$ is the elementary charge, and $Z_0$ is the characteristic impedance for the transmission line. The cosine function in $H_{\rm int}$ reflects the presence of a mirror giving an open boundary condition at $x = 0$.

Using the standard procedure of eliminating the photonic degrees of freedom under the Born-Markov approximation~\cite{Carmichael1999}, we obtain the interaction-picture master equation
\bea
\frac{d \rho}{dt} &=& i \sum_{n = 1}^N \delta_n \comm{\sigma_n^+ \sigma_n^-}{\rho} + i \sum_{n=1}^N \Omega_{\rm p}^n \cos(k_{\rm p} x_n) \comm{\sigma_n^x}{\rho} - i \sum_{n \neq m = 1}^N \mleft( \Delta_{nm}^+ - i \Gamma_{nm}^- \mright) \comm{\sigma_n^+ \sigma_m^-}{\rho} \nn \\
&&+ \sum_{n, m = 1}^N \mleft( \Gamma_{nm}^+ + i \Delta_{nm}^- \mright) \mleft( 2 \sigma_m^- \rho \sigma_n^+ - \sigma_n^+ \sigma_m^- \rho - \rho \sigma_n^+ \sigma_m^- \mright) + \sum_{n=1}^N \gamma_{\phi_n} \mleft( 2 \sigma_n^+ \sigma_n^- \rho \sigma_n^+ \sigma_n^- - \sigma_n^+ \sigma_n^- \rho - \rho \sigma_n^+ \sigma_n^- \mright),
\label{eq:ME}
\eea
where the qubit-qubit interaction is determined by
\bea
\Gamma_{nm}^+ &=& \frac{\gamma_{nm} + \gamma_{mn}}{2}, \\
\Gamma_{nm}^- &=& \frac{\gamma_{nm} - \gamma_{mn}}{2}, \\
\Delta_{nm}^+ &=& \frac{\Delta_{nm} + \Delta_{mn}}{2}, \\
\Delta_{nm}^- &=& \frac{\Delta_{nm} - \Delta_{mn}}{2},
\eea
with
\bea
\gamma_{nm} &=& \frac{\pi \alpha_{nm} \omega_{10}^m}{2} \mleft\{ \cos \mleft( k_m \mleft[ x_n + x_m \mright] \mright) + \cos \mleft( k_m \abs{x_n - x_m} \mright) \mright\}, \\
\Delta_{nm} &=& \frac{\pi \alpha_{nm} \omega_{10}^m}{2} \mleft\{ \sin \mleft( k_m \mleft[ x_n + x_m \mright] \mright) + \sin \mleft( k_m \abs{x_n - x_m} \mright) \mright\}, \\
\alpha_{nm} &=& \frac{2 \beta_n \beta_m e^2 Z_0}{\hbar \pi} \mleft( \frac{E_{\rm J}^{(n)}}{8 E_{\rm C}^{(n)}} \mright)^{1/4} \mleft( \frac{E_{\rm J}^{(m)}}{8 E_{\rm C}^{(m)}} \mright)^{1/4}.
\eea
In these expressions, the subscripts $n$ and $m$ refer to qubits $n$ and $m$; in general, these indices are not interchangeable in terms where they occur together if the two qubits they refer to are non-identical. The first term in \eqref{eq:ME} is the Hamiltonian for the individual qubits. Here, we have absorbed single-qubit Lamb shifts into the detuning $\delta_n$ between the frequency of qubit $n$ and the frequency $\omega_{\rm p}$ of a probe field:
\be
\delta_n = \omega_{\rm p} - \omega_n - \Delta_{nn}.
\ee
The second term in \eqref{eq:ME} is the Hamiltonian showing qubit $n$ is driven by the probe field, which is characterized by the Rabi frequency
\be
\hbar \Omega_{\rm p}^n = 2 \sqrt{2} e \beta_n \mleft( \frac{E_{\rm J}^{(n)}}{8 E_{\rm C}^{(n)}} \mright)^{1/4} V_0,
\label{eq:RabiFreq}
\ee
where $V_0$ is the input voltage. The third term in \eqref{eq:ME} is the qubit-qubit interaction that gives rise to the collective Lamb shift. The fourth term in \eqref{eq:ME} describes individual and collective relaxation processes for the qubits. We note that the individual bare decay rate for qubit $n$ is given by
\be
\gamma_n = \pi g_n^2 \mleft( \omega_{10}^n \mright) = \pi \alpha_{nn} \omega_{10}^n.
\ee
Finally, the fifth term in \eqref{eq:ME} describes pure dephasing. The pure dephasing rate of qubit $n$ is $\gamma_{\phi_n}$.


\section{Reflection coefficient}

In this section, we summarize the calculation for obtaining the reflection coefficient 
\be
r \equiv \abs{V_{\rm out} / V_{\rm in}}
\ee
from the qubits in the semi-infinite transmission line for an input voltage $V_{\rm in}$. The output voltage is given by
\be
V_{\rm out} (x, t) = V_{\rm in} (x, t) + V_{\rm s} (x, t),
\ee
where the scattered signal is
\be
V_{\rm s} (x, t) = - i \sqrt{\frac{\hbar Z_0}{4 \pi}} \int_0^\infty d \omega \sqrt{\omega} a_\omega (t) e^{i k x}.
\ee
Here, the photonic operator $a_\omega (t) = \tilde{a}_\omega (t) e^{-i \omega t}$ can be expressed in terms of the slowly varying amplitude 
\be
\tilde{a}_\omega (t) = - \sum_{n = 1}^N g_n (\omega) \int_0^t \tilde{\sigma}_n^- (s) e^{i (\omega - \omega_n) s} ds,
\ee
with $\tilde{\sigma}_n^- (t) = \sigma_n (t) e^{i \omega_{10}^n t}$. Substituting $\tilde{a}_\omega$ into $V_{\rm s}$ and performing the integration, we obtain
\be
V_{\rm s} (x, t) = i \sum_{n = 1}^N \sqrt{2} e \beta_n Z_0 \omega_n \mleft( \frac{E_{\rm J}^{(n)}}{8 E_{\rm C}^{(n)}} \mright)^{1/4} \cos \mleft( k_{\rm p} x_n \mright) \tilde{\sigma}_n^- (t).
\ee
Since the input signal $V_{\rm in}$ is connected to the Rabi frequency of the pumping field through \eqref{eq:RabiFreq} by taking $n = N$, we immediately obtain
\be
r = \abs{1 + i \sum_{m} \frac{4 \eta_{Nm} \gamma_m}{\Omega_{\rm p}^N} \cos \mleft( k_{\rm p} x_m \mright) \expec{\sigma_m^-}},
\ee
with 
\be
\eta_{Nm} = \frac{\beta_N}{\beta_m} \mleft(\frac{E_{\rm J}^{(N)} E_{\rm C}^{(m)}}{E_{\rm J}^{(m)} E_{\rm C}^{(N)}} \mright)^{1/4}.
\ee
The reflection coefficient can then be computed numerically by evolving the master equation in \eqref{eq:ME}.


\section{Full spectroscopy}

\begin{figure}
\includegraphics[width=\linewidth]{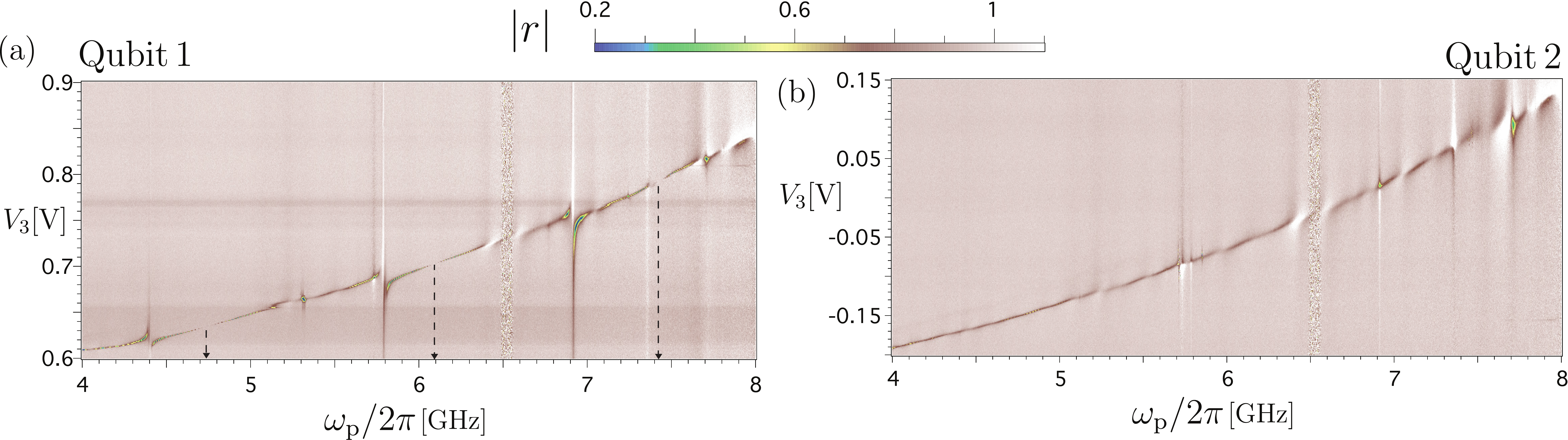}
\caption{Amplitude reflection coefficient $\abs{r}$ as a function of probe frequency $\omega_{\rm p}$ and a magnetic flux tuning the qubit frequencies for the full bandwidth of our measurement setup. In each measurement, the other qubit is detuned far from resonance.
(a) The data for Qubit 1, which is located away from the mirror. The dashed arrows indicate frequencies where the response shows that Qubit 1 sits at a node for the electromagnetic field in the transmission line. The marked frequencies are $f_1 = \unit[4.745]{GHz}$, $f_2 = \unit[6.094]{GHz}$, and $f_3 = \unit[7.414]{GHz}$; they correspond to $L = 7 \lambda_1 / 4$, $L = 9 \lambda_2 / 4$, and $L = 11 \lambda_3 / 4$, respectively. Knowing that $L = \unit[33]{mm}$, this lets us calculate the speed of light in the transmission line. We find $v = f_1 \lambda_1 = \unit[0.8948 \times 10^8]{m/s} \approx f_2 \lambda_2 \approx f_3 \lambda_3$.
(b) The data for Qubit 2, which is located right by the mirror.
\label{fig:FullBandwidthSpectroscopy}}
\end{figure}

In this section, we present the full data from the single-qubit spectroscopy, part of which was shown in Fig.~2 in the main text. Figure \ref{fig:FullBandwidthSpectroscopy} shows the amplitude reflection coefficient $\abs{r}$ as a function of probe frequency $\omega_{\rm p}$ and qubit frequency in the full range $\unit[4-8]{GHz}$, which is the bandwidth of the cryogenic low-noise amplifier in our experimental setup. As explained in the caption, we use this data to extract the speed of light in the transmission line.

In each of our transmon qubits, two capacitively shunted Josephson junctions form a SQUID loop. The external flux $\Phi$ through this loop affects the transition energy of the qubit~\cite{Koch2007}:
\be
\hbar \omega_{10} (\Phi) \approx \sqrt{8 E_{\rm J} (\Phi) E_{\rm C}} - E_{\rm C}.
\ee
The transition energy is determined by the charging energy $E_{\rm C} = e^2 / 2 C_\Sigma$ and the Josephson energy
\be
E_{\rm J} (\Phi) = E_{\rm J} \abs{\cos \mleft( \pi \Phi / \Phi_0 \mright)},
\ee
where $\Phi_0 = h / 2e$ is the magnetic flux quantum. The Josephson energy can be tuned from its maximum value $E_{\rm J}$ by the external flux $\Phi$ via a magnetic coil or local flux line.

\section{Additional information for figures in the main text}

For completeness, we here present the parameters extracted from fitting the linecuts in the single-tone spectroscopy shown in Fig.~2(e) and (f) in the main text. These parameters are given in \tabref{tab:ParametersSingleTone}.

\begin{table}
\centering
\begin{tabular}{| c |c | c | c | c | }
\hline
Qubit (Bias) & $\omega_{10} / 2 \pi$ [GHz] & $\Gamma_1 / 2 \pi$ [MHz] & $ \gamma_\phi / 2 \pi$ [MHz] & $\gamma / 2 \pi$ [MHz] \\
\hline
Q1 (A) & $4.697$ & $0.3$ & $2.1$ & $2.25$ \\
Q1 (B) & $5.01$ & $8$ & $1.7$ & $5.7$ \\
Q2 (C) & $4.692$ & $21$ & $2.15$ & $12.65$ \\
Q2 (D) & $5.014$ & $21$ & $2$ & $12.5$ \\
\hline
\end{tabular}
\caption{Extracted parameters from the linecuts $A$-$D$ in Fig.~2(e) and (f). The fit to theory is performed following Ref.~\cite{Hoi2013a}.
\label{tab:ParametersSingleTone}}
\end{table}

\begin{figure}
\includegraphics[width=0.5\linewidth]{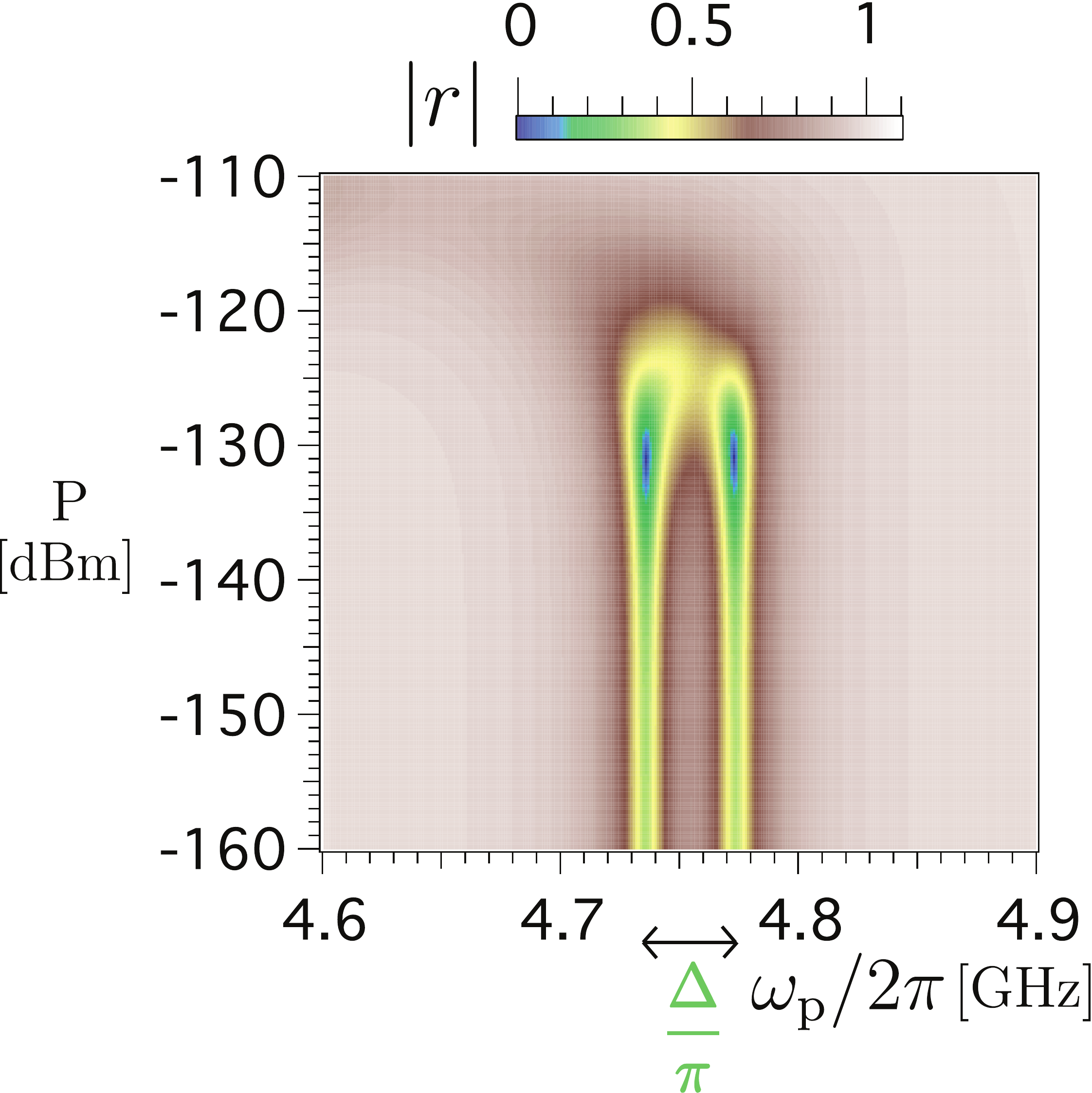}
\caption{Theoretical simulation of the results in Fig.~4(a) in the main text. The simulation uses parameter values extracted in earlier measurements given in the main text.
\label{fig:PowerDependenceTheory}}
\end{figure}

Finally, we also provide the theoretical simulation of the experimental results presented in Fig.~4(a) in the main text. These simulations are shown in \figref{fig:PowerDependenceTheory}.

\bibliography{LambShiftRefs}